\documentstyle[aps,prl]{revtex}
\input psfig.sty
\begin{document}
\twocolumn

\title{Watching a superfluid untwist itself: Recurrence of Rabi
oscillations in a Bose-Einstein condensate}
\author{M.R. Matthews, B.P. Anderson, P.C. Haljan, D.S. Hall,
 M.J. Holland, J.E Williams, C.E. Wieman, E.A. Cornell\cite{qpdNIST}}
\address{JILA, National
  Institute of Standards and Technology and Department of Physics,\\
  University of Colorado, Boulder, Colorado 80309-0440}
\date{June 16,1999}
\maketitle
\begin{abstract}
The order parameter of a condensate with two internal states can continuously
distort in such a way as to remove twists that have been imposed along its
length. We observe this effect experimentally in the collapse and recurrence of
Rabi oscillations in a magnetically trapped, two-component Bose-Einstein
condensate of $^{87}$Rb.
\end{abstract}
\par
Quantization and persistence of current in superconductors and
superfluids can be understood in terms of the topology of the
order parameter. Current arises from a gradient in the phase of
the order parameter. Quantization of flow around a closed path is
a consequence of the requirement that the order parameter be
single-valued; metastability, or ``persistence", arises from the
fact the number of phase windings (the multiple of $2\pi$ by which
the phase changes) around the path can be changed only by forcing
the amplitude of the order parameter to zero at some point.  If
the energy this requires exceeds that available from thermal
excitations, then the current will be immune to viscous damping.
This familiar argument relies, however, on the order parameter's
belonging to a very simple rotation group. The order parameter in
superfluid $^4$He, for example, is a single complex number. Its
phase can be thought of as a point lying somewhere on a circle,
and  is subject to the topological constraints mentioned above.
The order parameter of a more complicated superfluid, on the other
hand, will in general be capable of ridding itself of unwanted
kinetic energy by moving continuously through a higher-dimensional
order-parameter space in such a way as to reduce, even to zero,
its winding number. Presumably this ability will reduce a
superfluid's critical velocity; in the limit that the
order-parameter space is fully symmetric, the critical velocity
may even vanish \cite{Bhattacharyya1977}.
\par
In this paper, we discuss experiments on a gas-phase Bose-Einstein condensate
with two internal levels \cite{othermultiBEC}. This is equivalent to a spin-1/2
fluid: the order parameter has SU(2) rotation properties. A differential torque
across the sample is applied to the order parameter so that with time it
becomes increasingly twisted. Eventually the sample distorts through SU(2)
space so that the steadily applied torque now has the effect of untwisting the
order parameter, which returns nearly to its unperturbed condition. The pattern
of twisting and then untwisting is manifested experimentally as a washing-out
followed by a recurrence of an extended series of oscillations in the
population between the spin states. Related behavior has been previously
observed in A-phase $ ^3$He \cite{Paulson1976}; a major difference in this work
is that we can directly observe components of the order parameter with temporal
and spatial resolution.
\par
Magnetically confined $ ^{87}$Rb can exist in a superposition of
two internal states, known as \protect{$\left | 1 \right >$} and
\protect{$\left | 2 \right >$}  \cite{thestates}. The two internal
states are separated by the relatively large $ ^{87}$Rb hyperfine
energy, but in the presence of a  near-resonant coupling field the
states appear, in the rotating frame, to be nearly degenerate. The
condensate can then dynamically convert between internal states.
The order parameter for the condensate is the pair of complex
field amplitudes $\Phi_1$ and $\Phi_2$ of states \protect{$\left |
1 \right >$} and \protect{$\left | 2 \right >$}. Evolution of
these fields is governed by a pair of coupled Gross-Pitaevskii
equations which model the coupling drive, the external confining
potential, kinetic energy effects and mean-field interactions
\cite{Cornell1998a,Hall1998b,Dum1998a}. The SU(2) nature of the
order parameter ($\Phi_1$,$\Phi_2$) is more evident if we write

\begin{mathletters}
\begin{equation}
 \Phi_1 = \cos(\theta /2) e^{-i\phi/2} n_t^{1/2} e^{i\alpha}\label{su2a}
\end{equation}
and
\begin{equation}
 \Phi_2 = \sin(\theta /2) e^{i\phi/2} n_t^{1/2} e^{i\alpha}\label{su2b}
\end{equation}
\end{mathletters}

where $\theta, \phi, n_t,$ and $\alpha$ are purely real functions
of space and time.  $\theta$ and $\phi$ give the relative
amplitude and phase of the two internal components, and may be
thought of respectively as the polar and azimuthal angles of a
vector whose tip lies on a sphere in SU(2) space. The total
density and mean phase, $n_t$ and $\alpha$ respectively, remain
relatively constant \cite{Hall1998a} during the condensate
evolution described in this paper.
\par
The apparatus has been previously described
\cite{Cornell1998a,Hall1998b,Matthews1998a}. The starting point for the
measurements is a magnetically confined cloud of $\sim8 \times 10^5$
evaporatively cooled, Bose-Einstein-condensed ${}^{87}\mathrm{Rb}$ atoms near
zero temperature. The combined gravitational and magnetic potentials
\cite{Petrich1995a} yield an axially symmetric, harmonic confining potential
$V_1$ ($V_2$) for particles in the \protect{$\left |  1 \right >$}
(\protect{$\left |  2 \right >$}) state, in which the aspect ratio of the axial
oscillation frequency in the trap to the radial frequency $\omega_z/\omega_r$
can be varied from 2.8 to 0.95 \cite{Ensher98}. $V_1$ and $V_2$ are nearly
identical but can optionally be spatially offset a distance $z_0$ in the axial
direction \cite{Halloffset}. The coupling field has a detuning $\delta$ from
the local \protect{$\left |  1 \right
>$} to \protect{$\left |  2 \right >$} resonance. If $z_0$ is
nonzero, $\delta$ depends on the axial position $z$, with
$\delta(z)-\delta(z=0)$ linear in $z$ and in $z_0$. The strength, characterized
by the Rabi frequency $\Omega$, of the coupling field also varies with an axial
gradient \cite{twophoton}.

\par
We are able to measure the population of both spin states
nondestructively using phase-contrast microscopy
\cite{Zernike1953,Andrews1996}. We tune the probe laser between
the resonant optical frequencies for the \protect{$\left | 1
\right
>$} and \protect{$\left | 2 \right >$} states. Since the probe
detuning has opposite sign for the two states, the resulting phase
shift imposed on the probe light has opposite sign, such that the
\protect{$\left | 1 \right >$} atoms appear white and the
\protect{$\left | 2 \right >$} atoms appear black against a gray
background on the CCD array.  We can acquire multiple,
nondestructive images of the spatial distribution of the
\protect{$\left | 1 \right>$} and \protect{$\left | 2 \right >$
atoms at various discrete moments in time, or we can acquire a
quasi-continuous time record (streak image) of the difference of
the populations in the \protect{$\left | 1 \right>$} and
\protect{$\left | 2 \right >$ states, integrated across the
spatial extent of the cloud.

\par
The effect of the coupling drive is to induce a precession of the
order parameter at the local effective Rabi frequency $\Omega_{\rm
eff}(z)\equiv(\Omega(z)^2+\delta(z)^2)^{1/2}$.   In a preliminary
experiment, we chose parameters so as to make $\Omega_{\rm eff}$
nearly uniform, with $\omega_z=2\pi \times 63$Hz, $\omega_r=2\pi
\times 23$ Hz, $\Omega\simeq2\pi \times 340$ Hz and
$\delta(z)\simeq0$. A condensate at near-zero temperature was
prepared in the pure \protect{$\left | 1 \right>$} state. The
coupling drive was then turned on suddenly, inducing an extended
series of oscillations of the total population from the
\protect{$\left | 1 \right>$} to the \protect{$\left | 2 \right >$
state (``Rabi oscillations") \ [Fig. \ref{fig1}]. The robustness
of the Rabi oscillations is proof that our imaging does not
significantly perturb the quantum phase of the sample
\cite{freqshift}(population transfer via Rabi oscillations is
phase-sensitive).

\par
If there is an axial gradient to $\Omega_{\rm eff}$, then a
relative torque is applied to the order parameter across the
condensate, which can cause a twist to develop along the axial
direction. If we naively model the sample as a collection of
individual atoms, each held fixed at its respective location, then
the order parameter at each point in space rotates independently
at the local effective Rabi frequency, $\Omega_{\rm eff}$.  In
Fig. \ref{arrowfigs}(a) we see the implications of the
``fixed-atom" model for $\Omega=2\pi\times700$ Hz and
$\delta(z)=2\pi\times(100+14z)$Hz with $z$ in microns: a twist
develops in the order parameter which leads to a washing out of
the Rabi oscillations \ [Fig. \ref{arrowfigs}(c)]. In contrast,
the kinetic energy provides stiffness for a true condensate.
Simulation of the condensate \cite{simulation} shows that, for
early times, the order parameter begins twisting as in the
``fixed-atom" model but the twisting process self-limits about 40
ms into the simulation. At this point there is nearly a full
winding across the condensate. Thereafter, though the two ends of
the order parameter continue to twist with respect to one another,
the order parameter has been sufficiently wrapped around the SU(2)
sphere that the effect of further torque is to return the
condensate close to its unperturbed condition \ [Fig.
\ref{arrowfigs}(d)]. The Rabi oscillations exhibit a corresponding
revival \ [Fig. \ref{arrowfigs}(e)]. The factor driving the
untwisting process is the increasing kinetic energy cost
associated with an increasing twist in the order parameter. For a
simple U(1) order parameter, continuously increasing the winding
ultimately results in a ``snap", in which the order parameter is
driven to zero and a discontinuous (and presumably dissipative
\cite{Packard}) process releases the excess windings. The revival
in the present case is made possible by the larger rotation space
available to a two-component cloud.

\par
Under experimental conditions similar to those of the simulations
in Fig. \ref{arrowfigs}, we have observed as many as three
complete cycles of Rabi-oscillation decay and recurrence. These
data appear in Ref. \cite{Williams1999}. In this paper we present
data that correspond to the case of a more vigorous twisting. We
increase the axial dimension of the condensate cloud by a factor
of four; the kinetic energy cost of twisting the condensate is
correspondingly lower, so that at the point in time when the
condensate is maximally distorted there are four windings across
the cloud. The parameters of the experiment were as follows:
$\omega_r=\omega_z=2 \pi \times 7.8$ Hz and mean $\Omega_{\rm eff}
= 2 \pi \times 225$Hz. There was a gradient in both $\delta$ and
$\Omega$ across the $54 \mu m$ axial extent of the cloud,
resulting in a $\sim2\pi\times60$Hz difference in $\Omega_{\rm
eff}$ from top to bottom of the condensate. The result of the
experiment is seen in Fig. \ref{fig3}. The observed recurrence of
the Rabi oscillations at $180$ ms \ [Fig. \ref{fig3}(a)], when
corrected for overall decay of the cloud, corresponds to 60
percent contrast. We find it remarkable that the distorted
order-parameter field seen in Fig. \ref{fig3}(b) at times 65 and
75 ms should find its own way back to a nearly uniform
configuration.

\par
The simulations qualitatively reproduce the integrated number and
state-specific density distributions observed in the experiments.
For large inhomogeneity in $\Omega_{\rm eff}$, however, the
simulations predict the development of small-scale spatial
structure not observed in the experiment.  The simulations contain
no dissipation, whereas finite-temperature damping may occur in
the experiment \cite{Fedichev1998a}.
\par
Heuristically, what value do we expect for the recurrence time
$t_{\rm recur}$ for the data in Fig. \ref{fig3}?  The difference
in $\Omega_{\rm eff}$ from the top to the bottom of the condensate
is about 60 Hz. From the data in Fig. \ref{fig3}(b) we see that
the recurrence occurs only after four windings have one-by-one
been twisted in and then twisted out of the condensate. A rough
estimate then would be $t_{\rm recur}= (4+4)/60$Hz$=133$ ms,
shorter than the observed value of 180 ms, but reasonable given
that edge effects have been neglected.
\par
An interesting theoretical challenge would be to develop simple
arguments that would allow an {\it a priori} prediction of the
spacing of the windings at the instant of maximum twist.  For
particularly strong torques, one might expect the total density to
be suppressed to zero along a plane transverse through the cloud,
so as to allow for discontinuous relaxation of the order
parameter.  Indeed we have seen such behavior in numerical
simulations.  Under what conditions should suppression of total
density, rather than continuous evolution through SU(2) space, be
the preferred mode of relieving accumulated stress?
\par
We have observed that the presence of the coupling drive need not result in
population oscillations. For any given frequency and coupling strength, there
are two steady-state solutions which are completely analogous to the
dressed-states solutions of the single-atom problem \cite{Ballagh1999}. We have
been able experimentally to put the condensate in such states via an adiabatic
process: the strength of the drive is increased gradually from zero, and the
frequency (initially far detuned) is gradually ramped onto resonance. The
resulting ``dressed condensate" is extremely stable--- after the ramp is
complete, the cloud remains motionless in a near-equal superposition of the two
bare states. The twisting experiment discussed in this paper can be thought of
as the evolution of a highly nonlinear superposition of the two dressed states.
It would be interesting to work in the opposite limit and explore the spectrum
of {\it small}-amplitude excitations on a dressed state. Analogies to (i) the
transverse zero-sound in He$^3$ B-phase \cite{Lee} and (ii) spin-waves in
spin-polarized atomic hydrogen \cite{Bigelow1989a} should also be studied
\cite{spin}.
\par
The ability to fundamentally alter the topological properties of a condensate
has already proven useful. When the coupling drive is turned off, the SU(2)
properties vanish, and states \protect{$\left | 1 \right >$} and
\protect{$\left | 2 \right >$} become distinct species, each forced to live in
its separate U(1) space. Figure \ref{fig3} illustrates how this can be used to
change $\Phi_1(z)$ in a controlled manner. With a variation
\cite{Williamsvortex} on this technique we have created a vortex-state
condensate \cite{vortex}.
\par
We acknowledge funding from the ONR and the NSF, and useful
conversations with Jason Ho and Seamus Davis.

\begin{figure}[p]
\begin{center}
\psfig{figure=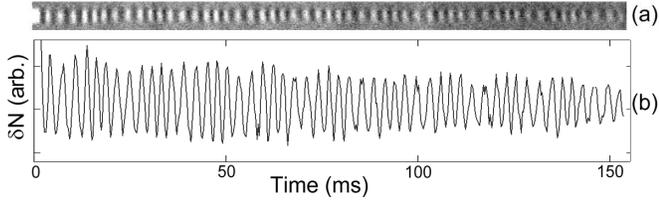,width=1\linewidth,clip=}
\end{center}
\caption{(a) With the trap parameters adjusted for high spatial
uniformity in $\Omega_{\rm eff}$, we drive the coupling transition
and record a streak-camera image of 60 Rabi oscillations between
the \protect{$\left | 1 \right >$} (white) and \protect{$\left | 2
\right >$} (black) states. The vertical dimension of the figure is
80 $\mu$m. (b) The value of $\delta N$, the total number of atoms
in \protect{$\left | 2 \right
>$} minus the total in \protect{$\left |  1 \right >$}, is extracted from the
image in part (a). The contrast ratio remains near unity ---
observed loss of signal is due to overall shrinkage of the
condensate through collisional decay.} \label{fig1}
\end{figure}

\begin{figure}[p]
\begin{center}
\psfig{figure=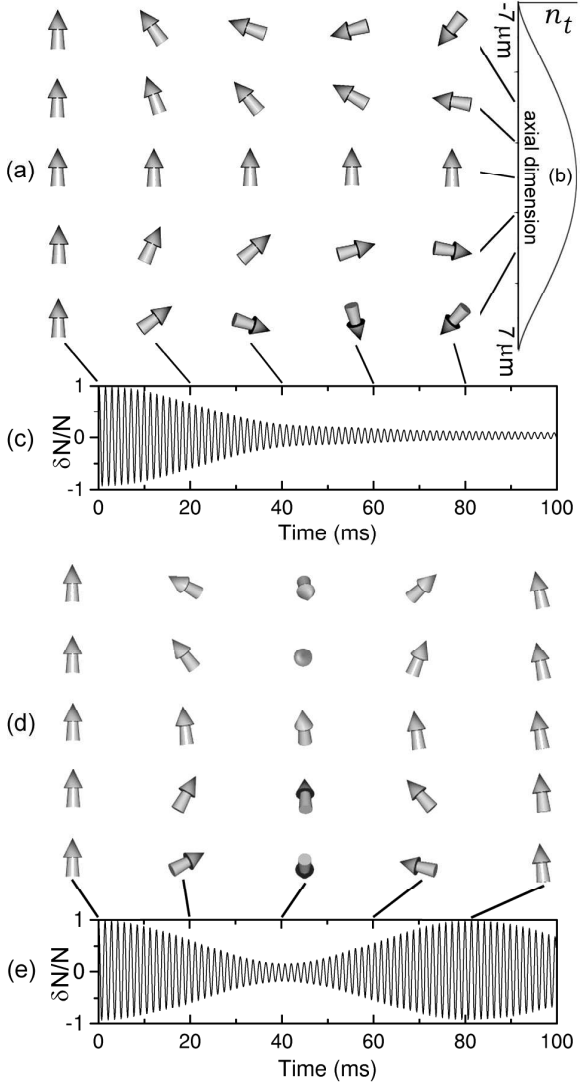,width=.88\linewidth,clip=}
\end{center}
\caption{(a) We represent the polar-vector order parameter as an
arrow in these simulations. The angle $\theta$ from the vertical
axis determines the relative population, and the azimuthal angle
$\phi$ is the relative phase of states \protect{$\left | 1 \right
>$} and \protect{$\left |  2 \right>$} (see Eqs. 1). Each column in
the arrow-array is at fixed time, and each row at fixed axial
location. $\hat{\Omega}$ is perpendicular to the plane of the
page, so that a uniform, on-resonance Rabi oscillation would
correspond to all the arrows rotating in unison, in the plane of
the image. The tips of all the arrows are (on the relatively fast
time scale of $\Omega_{\rm eff}$) tracing out circles nearly
parallel to the plane of the page (in our rotating-frame
representation, small excursions out of the page are a consequence
of finite detuning). In the figures, we ``strobe" the motion just
as the central arrow approaches vertical, to emphasize the more
slowly evolving ``textural" behavior. (b) The total density of the
condensate $n_t$ maintains a Thomas-Fermi distribution (integrated
through one dimension, as imaged) and changes only slightly during
the evolution of the cloud. (c) In a simple model of individual,
fixed atoms, a continuous inhomogeneity in $\Omega_{\rm eff}$ will
cause the Rabi oscillations in $\delta N$ to wash out. (d) When a
condensate is simulated [19], the kinetic energy causes the order
parameter to precess through the full SU(2) space, coming out of
the page to cast off the winding and thus reduce its kinetic
energy.  (e) The corresponding plot of $\delta N$ shows that when
the arrows are once more aligned, the Rabi oscillations recur.}
\label{arrowfigs}
\end{figure}

\pagebreak
\begin{figure}[p]
\begin{center}
\psfig{figure=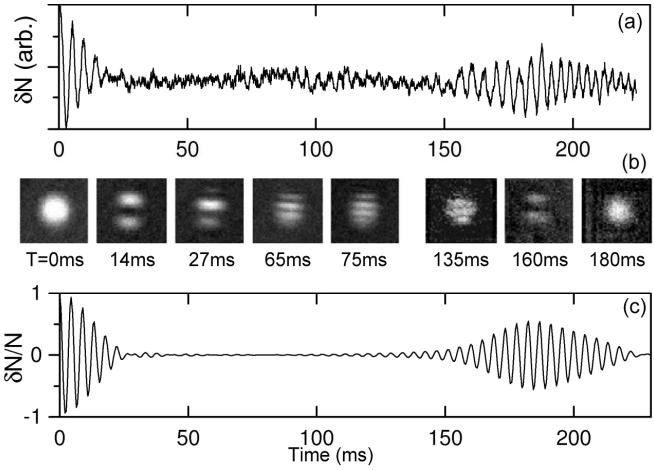,width=1\linewidth,clip=}
\end{center}
\caption {A condensate with large axial extent undergoes twisting.
(a) The streak camera data shows a rapid decay in the Rabi
oscillation in the integrated population difference, from full
contrast at t=0 to near zero by $t=20$ms. The oscillations recur
at 180 ms. (b) Individual phase-contrast images (at distinct
moments in time) of the spatial distribution of \protect{$\left |
1 \right
>$}-state atoms show that the spatially inhomogeneous Rabi
frequency is twisting the order parameter, cranking successively
more windings into the condensate until by $\sim 75$ ms four
distinct windings are visible. Further evolution results not in
more but in fewer windings until, at time $180$ ms, the order
parameter is once more uniform across the cloud. Each image block
is 100 $\mu$m on a side, and the probe laser is tuned much closer
to the \protect{$\left | 1 \right >$} state than to the
\protect{$\left | 2 \right
>$} state. (c) The numerical simulation reproduces the qualitative features of
the corresponding experimental plot (a). The simulation used
$\delta(z)=0$, $\Omega(z=0)=2\pi\times225$Hz and a
$2\pi\times60$Hz spread in $\Omega$ across the extent of the
condensate.}\label{fig3}
\end{figure}


\begin{references}
\bibitem{freqshift}
The probe beam does impose a small, reproducible shift on the
energy difference
  between the \protect{$\left | 1 \right >$} and \protect{$\left | 2 \right >$}
  states, a measurable shift for which we correct.
\bibitem{spin}
Obviously, oscillatory behaviour in coupled spin systems is a recurring theme.
At present it is unclear to us how deep the analogy runs.
\end{references}
\end{document}